\documentclass[11pt]{article}
\usepackage{graphicx}
\usepackage{subfigure}
\newcommand{\be}{\begin{equation}}
\newcommand{\ee}{\end{equation}}

\newcommand{\p}{\partial}
\newcommand{\f}{\frac}
\newcommand{\ri}{\rightarrow}

\newcommand{\bt}{\beta}

\begin{document}
\begin{titlepage}
\begin{flushright}
NTUA- 4/01
\end{flushright}

\vskip2truecm

\begin{center}
\begin{large}
{\bf Multi--Layer Structure \\ in the Strongly Coupled 5D Abelian Higgs Model}

\end{large}
\vskip1truecm

P.~Dimopoulos$^{(a)}$\footnote{E-mail: pdimop@central.ntua.gr},
K.~Farakos$^{(a)}$\footnote{E-mail: kfarakos@central.ntua.gr},
and S.~Nicolis$^{(b)}\footnote{E-mail:
nicolis@celfi.phys.univ-tours.fr}
$

\vskip1truecm

{\sl $^{(a)}$ Physics Department, National Technical University\\
15780 Zografou Campus, Athens, Greece

\vskip0.5truecm

\vskip0.5truecm

$^{(b)}$ CNRS-Laboratoire de Math\'ematiques et Physique
Th\'eorique (UMR 6083)
\\ Universit\'e de Tours, Parc Grandmont, 37200 Tours,France}

\end{center}
\vskip1truecm

\begin{abstract}

\noindent We  explore the phase diagram of the five-dimensional
anisotropic Abelian Higgs model by
Monte Carlo simulations.
In particular, we study the transition between the confining phase
and the  four dimensional layered Higgs phase. We find that,
in a certain region of the lattice parameter space, this
transition can be first order and that each layer  moves into
the Higgs phase independently of the others (
decoupling of layers). As the Higgs couplings
vary, we find, using mean field techniques, that this
transition may probably become second order.

\end{abstract}
\end{titlepage}

\section{Introduction}
Coupling anisotropies in gauge theories may lead to fundamental changes
in their phase diagrams. Although
this yields problematic theories if applied to four-dimensional models,
due to breakdown of Lorentz invariance, higher-dimensional models
with anisotropic couplings may give rise to theories of physical interest.
This programme originated in 1984 by Fu and Nielsen \cite{funiel}
who considered a five-dimensional pure U(1) gauge theory on the lattice.
The main idea is that, for
certain  values of the couplings for the $n$ extra dimensions,
four-dimensional layers may be formed within the $(4+n)-$dimensional
space. The corresponding phase is called {\it layered } phase and one of
its main
characteristics is that exhibits confinement in the extra
dimensions. The U(1) higher-dimensional model has already been
studied to some
extent with lattice techniques (\cite{stam}, \cite{dim1}),
leading to the establishment of the existence
of a layered Coulomb phase. \footnote {
The existence of a layered phase in non-Abelian theories
at finite temperature \cite{china} has been considered as well.}

The possibility that physical
space--time is not four-dimensional has been broadly referred to in
the bibliography during the last eighty years. This
interesting idea has enjoyed a revival through works that use
extra dimensions to solve the hierarchy problem. A class of these
theories use a (4+n)-dimensional
space--time with n compactified dimensions; another class of models
considers {\em non--compact} extra dimensions. A well--known example
of this last case
is the Randall--Sundrum (RS) model \cite{rs}, where the
four-dimensional world is considered as a three--brane embedded in a
five-dimensional bulk. Furthermore in such models a
four--dimensional graviton exists and is localised within the three-brane.
The question which then arises is if there is any
possibility that other fields, such as gauge fields, fermions and
scalars are localised within a 3-brane.
The problem has been attacked analytically through the search
for localized four--dimensional
fields, where the equations of motion of the bulk fields which are
coupled to the background geometry are solved \cite{alex}. In this
approach, using perturbative tools, a massless photon always
appears propagating freely in five dimensions.
A first attempt of getting
evidence of gauge field localization on a brane considering the
non--perturbative features while using an RS action type for abelian gauge
field has been performed in \cite{dim1} by means of lattice techniques.

Theories on flat spacetimes are not, however, devoid of interest.
In this paper we continue to explore the phase structure of the
Abelian Higgs model in five dimensions with anisotropic couplings,
defining our model on the lattice.
In a previous paper\cite{dim2} we studied the phase diagram of this model
for {\em weak} gauge coupling in the four-dimensional subspace and found two
kinds of layers (or 3-branes): of Coulomb or Higgs type. In the present
paper we wish to explore the strongly coupled theory, motivated, in part, by
recent considerations that indicate that theories studied in ref.~\cite{alex}
are, generically, strongly coupled.

We find that it is, indeed,
 possible to tune  the lattice couplings in such a way
that  a phase transition from the five-dimensional
strong phase to a layered phase occurs.
By measuring several order parameters we find that the
layers are in the Higgs phase and they are separated from each other by a
confining force. Furthermore we show the precise way in which they
are created by studying in detail the phase transition.
In particular, we find that each layer emerges from the confining phase
independently from the others-characteristic of a strongly first order
phase transition with a very small correlation length in the extra direction.

In section 2 we write down the Abelian Higgs model in five
dimensions with anisotropic couplings. In section 3 we present the
Monte--Carlo results, we exhibit the phase structure and we
establish the existence of the layered Higgs phase. Also, using
mean field techniques, we confirm the Monte--Carlo results and we
give indication of how the situation changes as we vary the Higgs
lattice couplings.

\section{Formulation of the model}

The model under study is the Abelian Higgs model in the five-dimensional
space. Direction $\hat 5$ will be singled
out by couplings that will differ from the corresponding ones
in the remaining four directions.

We proceed with writing down the lattice action of the model.
\begin{eqnarray}
S= &\beta_{g}& \sum_x\sum_{1 \le \mu<\nu \le 4}(1-\cos F_{\mu \nu}(x))
+\beta_g^{\prime}\sum_x\sum_{1 \le \mu \le 4}(1-\cos F_{\mu 5}(x))
\nonumber \\
&+&\beta_{h}\sum _{x} {\rm Re} [4 \varphi^{*}(x)\varphi (x)
- \sum_{1 \le \mu \le 4} \varphi^{*}(x)U_{\hat \mu}(x) \varphi (x+\hat \mu)]
\nonumber \\
&+&\beta_{h}^{\prime} \sum _{x} {\rm Re} [\varphi^{*}(x)\varphi (x)
- \varphi^{*}(x)U_{\hat 5}(x) \varphi (x+\hat 5)]
\nonumber \\
&+&\sum _{x}[(1-2\beta_{R}-4 \beta_{h}- \beta_{h}^{\prime})\varphi^{*}(x)\varphi (x)
+\beta_{R}(\varphi ^*(x)\varphi (x))^2],
\label{compactaction}
\end{eqnarray}
where
$$
F_{\mu \nu}(x)=A_\mu(x)+A_\nu(x+\hat \mu)-A_\mu(x+\hat \nu)-A_\nu(x),
~~1 \le \mu<\nu \le 4,
$$
$$
F_{\mu 5}(x)=A_\mu(x)+A_5(x+\hat \mu)-A_\mu(x+\hat 5)-A_5(x)
~~1 \le \mu \le 4.
$$

We have allowed for different couplings in the various directions: the ones
pertaining to the fifth direction are primed to distinguish them from
the ``space-like" couplings. The fifth direction will also be called
``transverse" in the sequel.

The link variables
$U_{\hat \mu}(x)$ are defined as $e^{i \alpha_S {\overline A_S}}$ or
$e^{i \alpha_T {\overline A_T}}$ respectively, where
${\overline A_S},~~{\overline A_T}$ are the continuum fields and
$\alpha_S,~\alpha_T$ are the lattice spacings in the space-like and
the transverse-like dimensions respectively.
The lattice fields are $$A_S \equiv \alpha_S {\overline A_S},~~
A_T \equiv \alpha_T {\overline A_T}.$$
In addition, the scalar fields are also written in the polar form
$\varphi(x) = \rho(x) e^{i \chi(x)}.$ The order parameters that we will use are
the following:

\be
{\rm Space-like~Plaquette:~~~} P_S \equiv <\f{1}{6 N^5} \sum_x
\sum_{1 \le \mu<\nu \le 4} \cos F_{\mu \nu}(x)>
\ee
\be
{\rm Transverse-like~Plaquette:~~~} P_T \equiv <\f{1}{4 N^5}
\sum_x \sum_{1 \le \mu \le 4} \cos F_{\mu 5}(x)>
\ee
\be
{\rm Space-like~Link:~~~} L_S \equiv <\f{1}{4 N^5} \sum_x
\sum_{1 \le \mu \le 4} \cos(\chi(x+\hat \mu) +A_{\hat \mu}(x)-\chi(x)) >
\ee
\be
{\rm Transverse-like~Link:~~~} L_T \equiv <\f{1}{N^5} \sum_x
\cos(\chi(x+\hat 5) +A_{\hat 5}(x)-\chi(x))>
\ee
\be
{\rm Higgs~field~measure~squared:~~~} R^2 \equiv \f{1}{N^5} \sum_x \rho^2(x)
\ee
In the above equations $N$ is the linear dimension of a symmetric $N^5$
lattice.

When necessary we will use the order parameters $L_{S}$ and
$P_{S}$ defined on each space--like volume (layer) separately.

The na\"{\i}ve continuum limit of the lattice action \ref{compactaction}
may be obtained as follows (where an overbar is used for the
continuum fields):
$$
\varphi={\overline \varphi} \sqrt{ \f{2 a_S^2 a_T}{\bt_{h}} },
$$
$$
~~ A_\mu = a_S {\overline A_\mu}, ~~1 \le \mu \le 4,
$$
$$
~~A_5 = a_T {\overline A_5}.
$$
Then the transverse-like field strength $$F_{\mu5} \equiv A_\mu(x)+
A_5(x+\hat \mu)-A_\mu(x+\hat 5)-A_5(x)~~(1 \le \mu \le 4)$$
goes over to:
$$-a_S [a_T \p_5 {\overline A_\mu}(x)]+a_T [a_S \p_\mu {\overline A_5}(x)] =
a_S a_T (\p_\mu {\overline A_5}-\p_5 {\overline A_\mu}).$$
Thus $$F_{\mu5}^2 \ri a_S^2 a_T^2 {\overline F_{\mu5}^2},
~~1 \le \mu \le 4~~(F_{\mu 5} \equiv \p_\mu {\overline A_5}
-\p_5 {\overline A_\mu}).$$
The space-like field strength is treated in a very similar way
with the result:
$$F_{\mu\nu}^2 \ri a_S^4 {\overline F_{\mu\nu}^2},~~
{\overline F_{\mu\nu}} \equiv \p_\mu {\overline A_\nu}
-\p_\nu {\overline A_\mu},~~1 \le \mu < \nu \le 4.$$

This means that the transverse-like part of the pure gauge action
is rewritten in the form:
$$\f{1}{2} \f{{\bt_g}' a_T}{a_S^2} \sum a_S^4 a_T
[\sum_{1 \le \mu \le 4} {\overline F}_{\mu 5}^{2}]
\rightarrow \f{1}{2} \f{{\bt_g}' a_T}{a_S^2} \int d^5 x [\sum_{1 \le \mu \le 4} {\overline F}_{\mu 5}^{2}].$$
On the other hand the space--like part is:
$$\f{1}{2} \f{\bt_g}{a_T} \sum a_S^4 a_T
[\sum_{1 \le \mu <\nu \le 4} {\overline F}_{\mu \nu}^{2}]
\rightarrow \f{1}{2} \f{\bt_g}{a_T} \int d^5 x [\sum_{1 \le \mu < \nu \le 4}
{\overline F}_{\mu \nu}^{2}].$$
If we define
\be
\bt_{g} \equiv \f{a_T}{g_{S}^2},
~~{\bt_g}' \equiv \f{a_S^2}{g_{T}^{2} a_{T}},
\ee
the resulting continuum action reads:
$$\f{1}{2} \int d^5 x \left [ \f{1}{g_S^2} \sum_{1 \le \mu < \nu \le 4} {\overline F_{\mu\nu}}^2 +\frac{1}{g_{T}^2} \sum_{1 \le \mu \le 4} {\overline F_{\mu5}}^2 \right ]$$
Defining $\gamma_{g} \equiv (\frac{{\beta_g}' }{\beta_{g}})^{1/2}$ and using
the definitions of $\beta_{g}$, ${\beta_g}'$ we find that
$$\gamma_{g}= \frac{g_S}{g_T}\frac{a_S}{a_T}.$$
We denote by $\xi$ the important ratio
$\frac{a_S}{a_T}$ of the two lattice spacings (the {\it correlation anisotropy
parameter}) and finally derive the relation:
$$\gamma_{g}=\sqrt{ \f{{\beta_g}'}{\bt_g} }
= \frac{g_S}{g_T} \xi.$$
After rescaling the scalar fields, one may rewrite the scalar sector of the action in the
form:
\be
\int d^5 x [\sum_{1 \le \mu \le 4} |D_\mu {\overline \varphi}|^2
+\f{\gamma_\varphi^2}{\xi^2} |D_5 {\overline \varphi}|^2+
m^2 {\overline \varphi}^* {\overline \varphi}
+ \lambda ({\overline \varphi}^* {\overline \varphi})^2 ],
\label{contscal}
\ee
where $D_\mu \equiv \p_\mu-i {\overline A_\mu},~~1 \le \mu \le 5.$

We have used the notations:
$$ \gamma_\varphi \equiv \sqrt{ \f{{\beta_h}'}{\beta_{h}} },$$
$$ m^2 a{_S}^{2} \equiv \frac{2}{\beta_{h}}(1-2\beta_{R}
-4\beta_{h}-{\beta_h}'),~~\f{\lambda}{a_S}=\f{4\beta_R}{ \beta_{h}^2 \xi}.$$

If we choose a common value for the gauge coupling constants:
$g_S=g_T \equiv g,$ (so that $\gamma_g=\xi,$)
and assume that all the covariant derivatives
in equation (\ref{contscal}) have the same factor in front:
$\gamma_\phi=\xi$, the expression does not exhibit any anisotropy.
However, the  na\"{\i}vet\'{e} of this  approach will be  manifest by results
similar to Ref. \cite{stam,dim1,dim2} which indicate that the anisotropy
may survive in the continuum limit for a wide range of values of lattice
parameters both in flat as well as in warped (Randall-Sundrum type) spacetimes.

\section{Monte Carlo Results \\ and the Confining--Layered Transition}
For the simulations we use 5-hit Metropolis algotithm for the updating
of both the gauge and Higgs fields. In order to get better behaviour
we use a global radial algorithm and an overrelaxation
algorithm for the updating of the Higgs field.
We simulated the system for $4^{5}, 6^{5}, \mbox{and} \hspace{0.1cm} 8^{5}$
lattices. We made use, mainly, of the hysteresis loop method
to establish the phase diagram of the system.
When  necessary, in order to define more precisely the phase transition
points and study the order of the phase transition
we made long runs consisting up to 30000 measurements  at selected
points in the parameter space.

In whole work we set the four--dimensional gauge coupling
fixed at the value $\beta_{g}=0.5$ and let
the gauge coupling in the fifth direction run.
For some regions of the values of the two gauge couplings we had
a confining five--dimensional theory. A small value for the
Higgs coupling constant $\beta_{h}^{\prime}$
in the fifth dimension has been chosen
($\beta_{h}^{\prime}=0.001$) (we further discuss this
choice  in Section 3.2) and we used two values for the Higgs
self--coupling $\beta_{R}$  differing by one order of magnitude:
$\beta_{R}=0.01$ and $\beta_{R}=0.1$.
Thus the phase diagram has been found in the  $\beta_{g}^{\prime} - \beta
_{h}$ subspace.

\begin{figure}[!h]
\begin{center}
\includegraphics[scale=0.45]{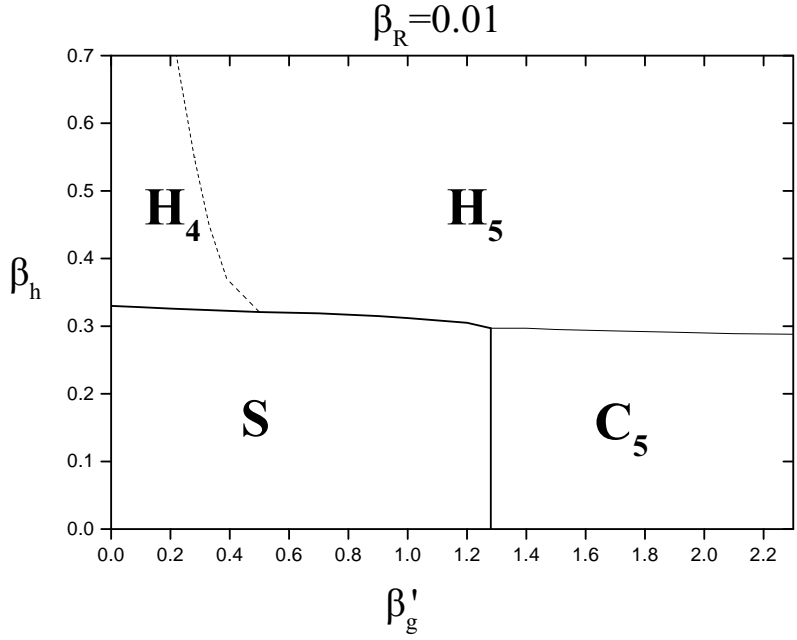}
\caption{Phase diagram for $\beta_{R}=0.01$}\label{Fig1}
\end{center}
\end{figure}

\begin{figure}[!h]
\begin{center}
\includegraphics[scale=0.45]{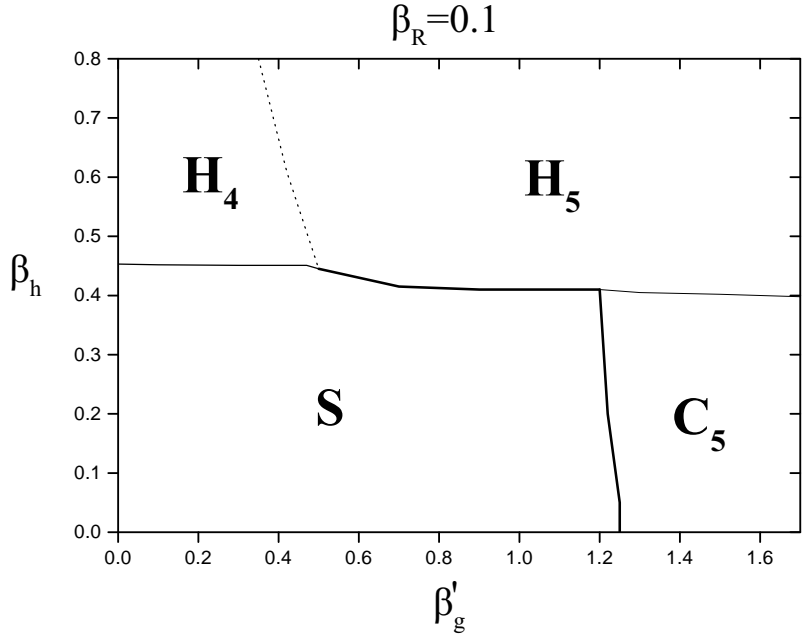}
\caption{Phase diagram for $\beta_{R}=0.1$}\label{Fig2}
\end{center}
\end{figure}
We study the behaviour of the system in terms of the order parameters defined in
Section 2. We proceed now with the presentation of the phase
structure.

Figures \ref{Fig1} and \ref{Fig2} show the phase diagrams of the 5--dimensional
Abelian Higgs model for the cases of two values of $\beta_{R}$
coupling, namely $\beta_{R}=0.01$ and $\beta_{R}=0.1$,
respectively.
The phase diagram for both cases has been explored in
$\beta_{g}^{\prime} - \beta_{h}$ subspace. In Table 1 we show the
values for the couplings  from which we deduced the phase
diagrams.
\begin{table}
\begin{tabular}{||c|cc|cc||}\hline
                 & \hspace*{2.8cm}$\beta_{R}=0.01$& & &\hspace*{-3.5cm}$\beta_{R}=0.1$ \\ \cline{2-5}
                 & $\beta_{g}^{\prime}$&$\beta_{h}$&
                 $\beta_{g}^{\prime}$&$\beta_{h}$ \\ \hline
$S - H_{4}$        &0.05, 0.10, 0.20, & & 0.10, 0.20, 0.25,&  \\
                   &0.40, 0.45 &&0.30, 0.40, 0.45, 0.46& \\ \hline
$S - H_{5}$       &0.50, 0.70, 0.80,&&0.47, 0.50, 0.70,& \\
                 &0.90, 1.00&&0.90, 1.00, 1.10& \\ \hline
$C_{5} - H_{5}$ &1.40, 1.50, 1.90&&1.20, 1.30, 1.50& \\
                &&&& \\ \hline
$H_{4} - H_{5}$  &&0.40, 0.50, 0.60&&0.50, 0.60, 0.80 \\
                  &&&& \\ \hline
$S - C_{5}$     & &0.05, 0.15, 0.20, &&0.10, 0.20, 0.30 \\
                &&0.25, 0.30&& \\ \hline
\end{tabular} \vspace{0.2cm}
\caption{The phase  which the phase diagrams given in Figures
\ref{Fig1} and \ref{Fig2} are based on.}
\end{table}
The two phase diagrams exhibit similar structure. There are four
different phases namely the strong phase (S), the Coulomb phase in
five dimensions ($C_{5}$) and the Higgs phases in four ($H_{4}$) and five
dimensions ($H_{5}$). More details about the nature of the
different phase transitions will be given later. The crucial
feature, however, is the existence of a phase transition between
the strong phase (eg. a confined
phase in five dimensions) and the $H_{4}$ phase --a phase of broken U(1)
symmetry in
four dimensions-- which gives rise to a constitution of a layered
phase with broken symmetry on each layer.
Let us give now  some representative results which lead
to the identification of the various phases of the model.
\vspace{0.2cm}

 {\bf  $\bullet$ \hspace{0.1cm} The $\mathbf{S - H_{4}}$ and
 $\mathbf{S - H_{5}}$ phase
transitions}\vspace{0.05cm}

\begin{figure}[!h]
\subfigure[]{\includegraphics[scale=0.3]{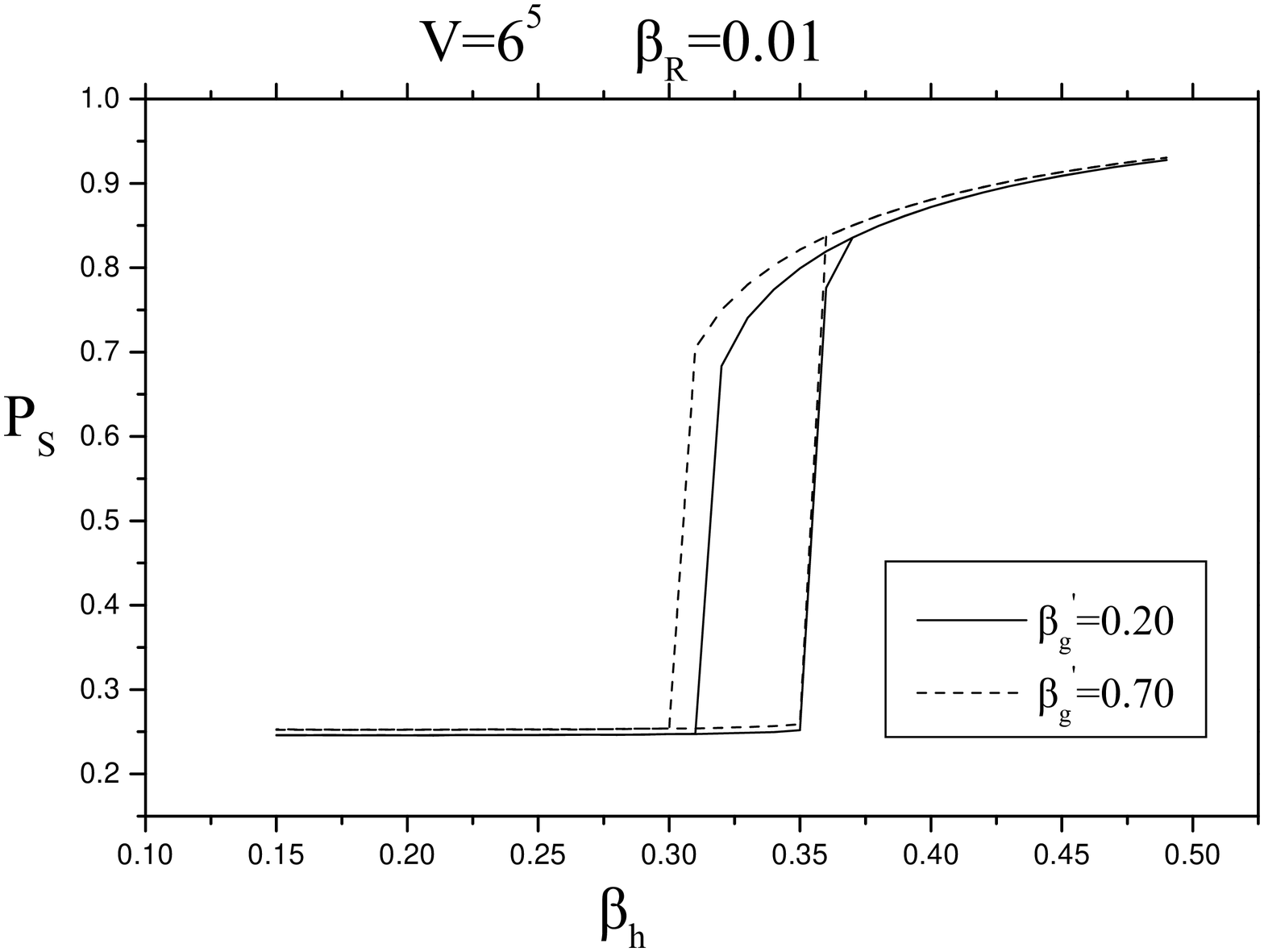}}
\subfigure[]{\includegraphics[scale=0.3]{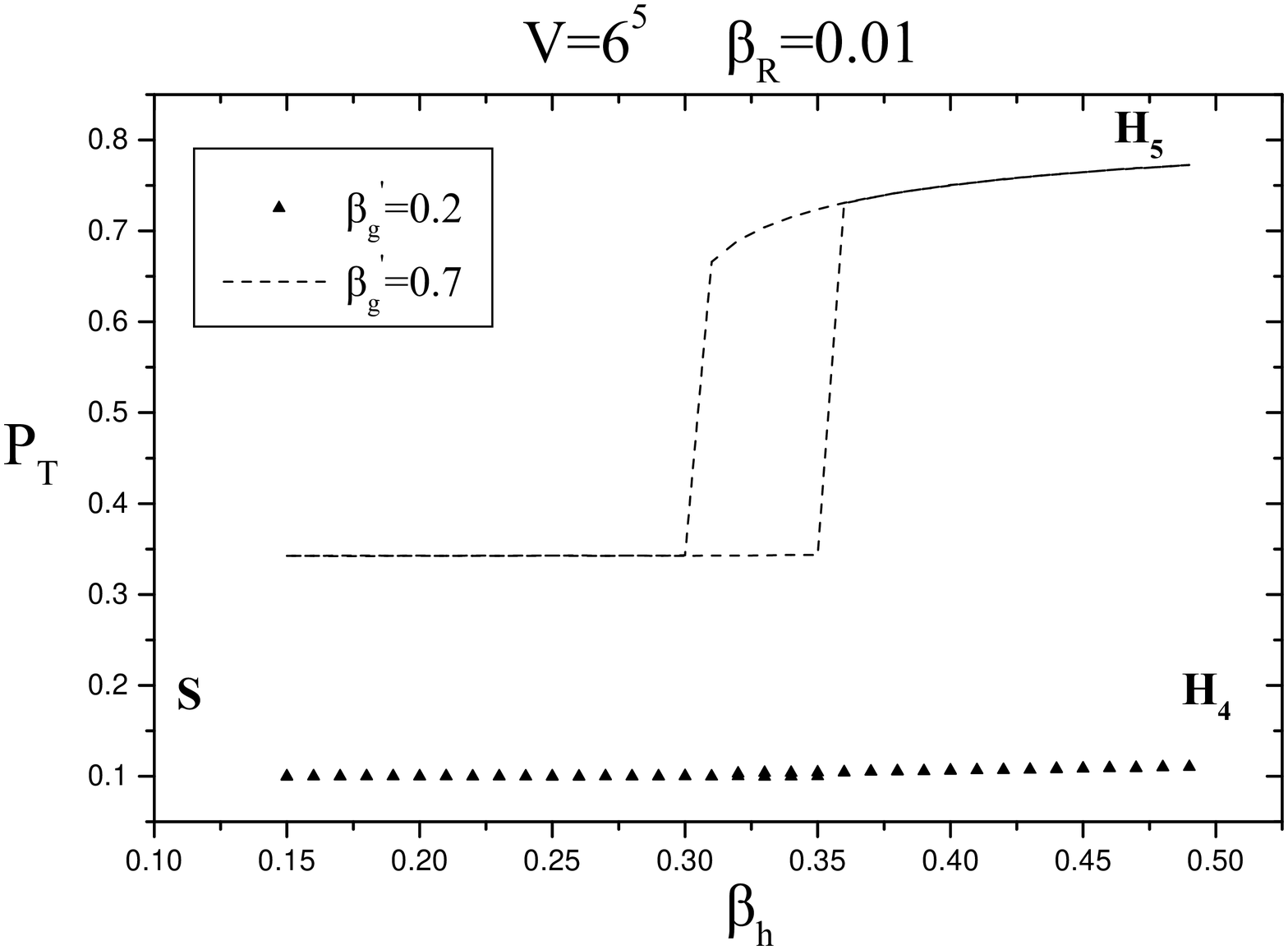}}
\caption{Hysteresis loop for  space--like (a) and transverse--like (b) plaquette
for two values of $\beta_{g}^{\prime}=0.2$ and 0.7.
}\label{Fig3}
\end{figure}

\noindent In Figure
\ref{Fig3}(a) , we present the hysteresis loops concerning the space--like
plaquettes, as $\beta_{h}$ runs, for $\beta_{R}=0.01$ and
for two values $\beta_{g}^{\prime}=0.2$ and
$\beta_{g}^{\prime}=0.7$. In Figure \ref{Fig3} (b) the
corresponding results for the transverse--like plaquettes are
depicted.

\begin{figure}[!h]
\begin{center}
\includegraphics[scale=0.3]{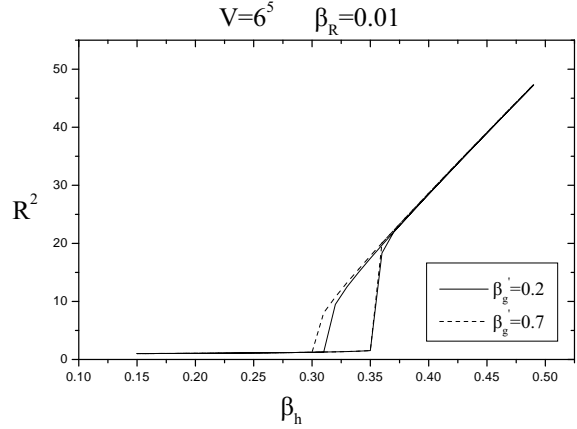}
\caption{Hysteresis loop for $R^2$, $\beta^{\prime}_{g}=0.2, 0.7$.}\label{Fig4}
\end{center}
\end{figure}

\begin{figure}[!h]
\begin{center}
\includegraphics[scale=0.3]{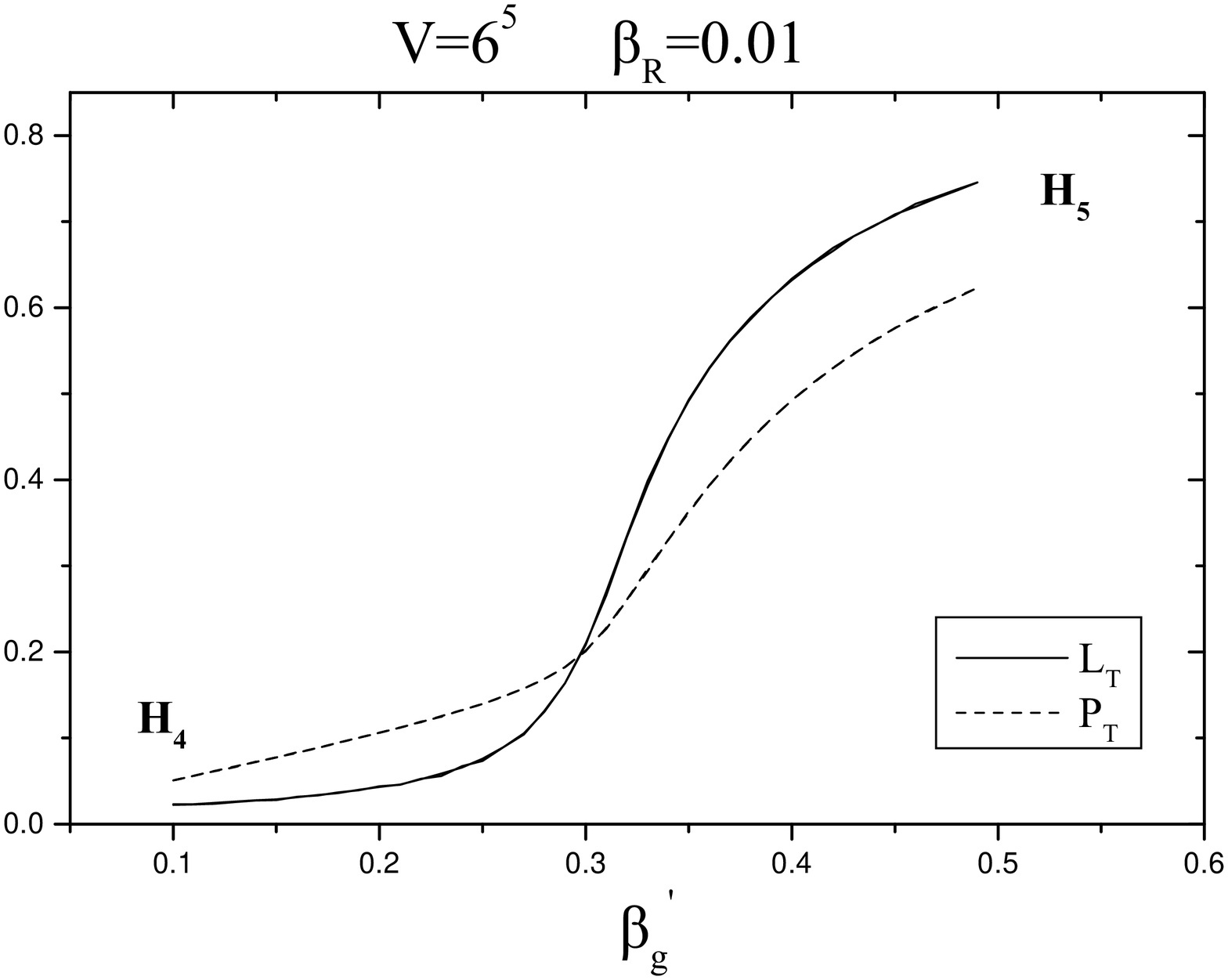}
\caption{Hysteresis loop analysis for $L_{T}$ and $P_{T}$ in the region between $H_{4}$ and
$H_{5}$ phases. The transition is not
showing formation of a loop and seems to be fairly smooth. }\label{Fig5}
\end{center}
\end{figure}
\noindent The behaviour of the space--like plaquettes and also the behaviour
of $R^2$, shown in Figure \ref{Fig4}, lead to the conclusion of a phase
transition between the five dimensional strong phase and a phase
with broken symmetry which probably is of first order by virtue
of the large hysteresis loop. In addition, the transverse--like
plaquette, $P_{T}$, for $\beta_{g}^{\prime}=0.2$ remains almost constant to a
small value (it equals the  value $\beta_{g}^{\prime}/2$, labeling the
Strong phase)
while the corresponding one
for $\beta_{g}^{\prime}=0.7$ increases with $\beta_{h}$ as a
phase transition occurs. This is a serious
indication that there are two different Higgs phases:
in particular one is a Higgs phase in four dimensions (with confining
behaviour along the fifth dimension) and the other is a five--dimensional
Higgs phase. \vspace{0.2cm}

{\bf $\bullet$ \hspace{0.1cm} The $\mathbf{H_{4} - H_{5}}$ phase transition} \vspace{0.05cm}

\noindent The fact that the fifth dimension is confining is made clearer
from the result of Figure \ref{Fig5}: we keep $\beta_{h}=0.40$ and
let $\beta_{g}^{\prime}$ run. The results depicted in this figure
correspond to the values of transverse--like link and
transverse--like plaquette which are small enough for small values
of $\beta_{g}^{\prime}$ and they increase as this coupling
parameter is running to larger values. It can be noticed that
there is no obvious hysteresis loop formed and  the system
passes from the $H_{4}$  (where the two order parameters take
small values) to the $H_{5}$ phase in a fairly smooth way.
\vspace{0.2cm}
\newpage

{\bf $\bullet$ \hspace{0.1cm} The $\mathbf{S - C_{5}}$ and $\mathbf{C_{5} - H_{5}}$
phase transitions} \vspace{0.05cm}

\begin{figure}[!h]
\begin{center}
\includegraphics[scale=0.3]{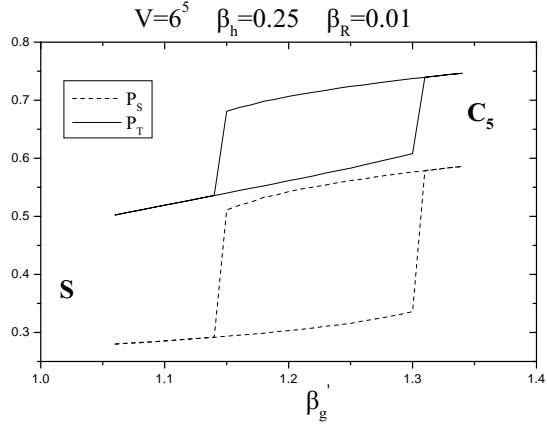}
\caption{Hysteresis loop for $P_{S}$ and $P_{T}$ for the $S - C_{5}$
phase transition. }\label{Fig6}
\end{center}
\end{figure}

\noindent The existence of the $C_{5}$ phase
is indicated in Figure \ref{Fig6}, which
contains the hysteresis loop for  $P_{S}$ and $P_{T}$
for running $\beta_{g}^{\prime}$. As it  can be easily seen, these two quantities pass
from a region where their values are almost half of their
corresponding gauge couplings (Strong phase) to a phase ($C_{5}$)where
their values tend to one. The large hysteresis loops indicate a
first order phase transition for $S - C_{5}$.

\begin{figure}[!h]
\subfigure[]{\includegraphics[scale=0.3]{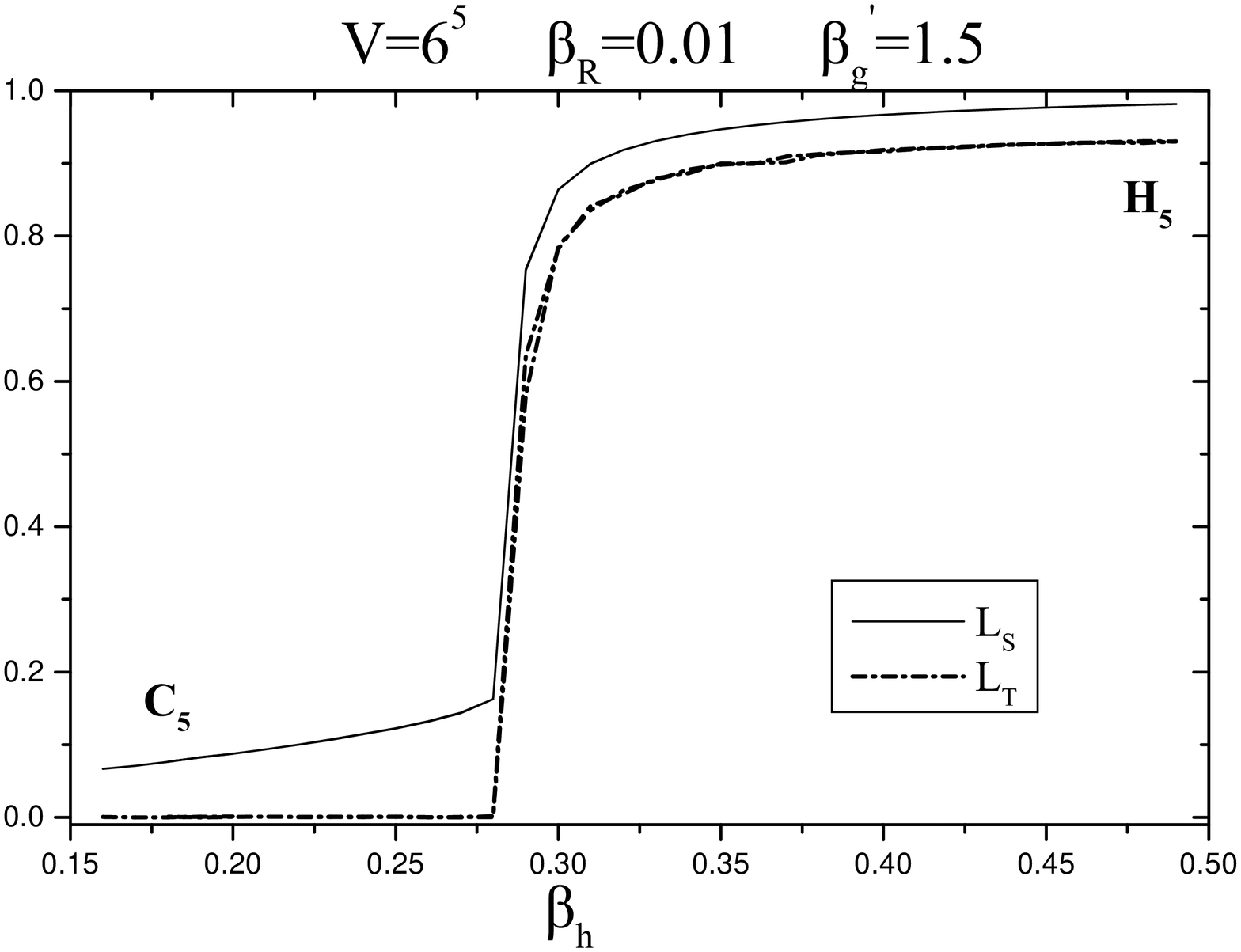}}
\subfigure[]{\includegraphics[scale=0.3]{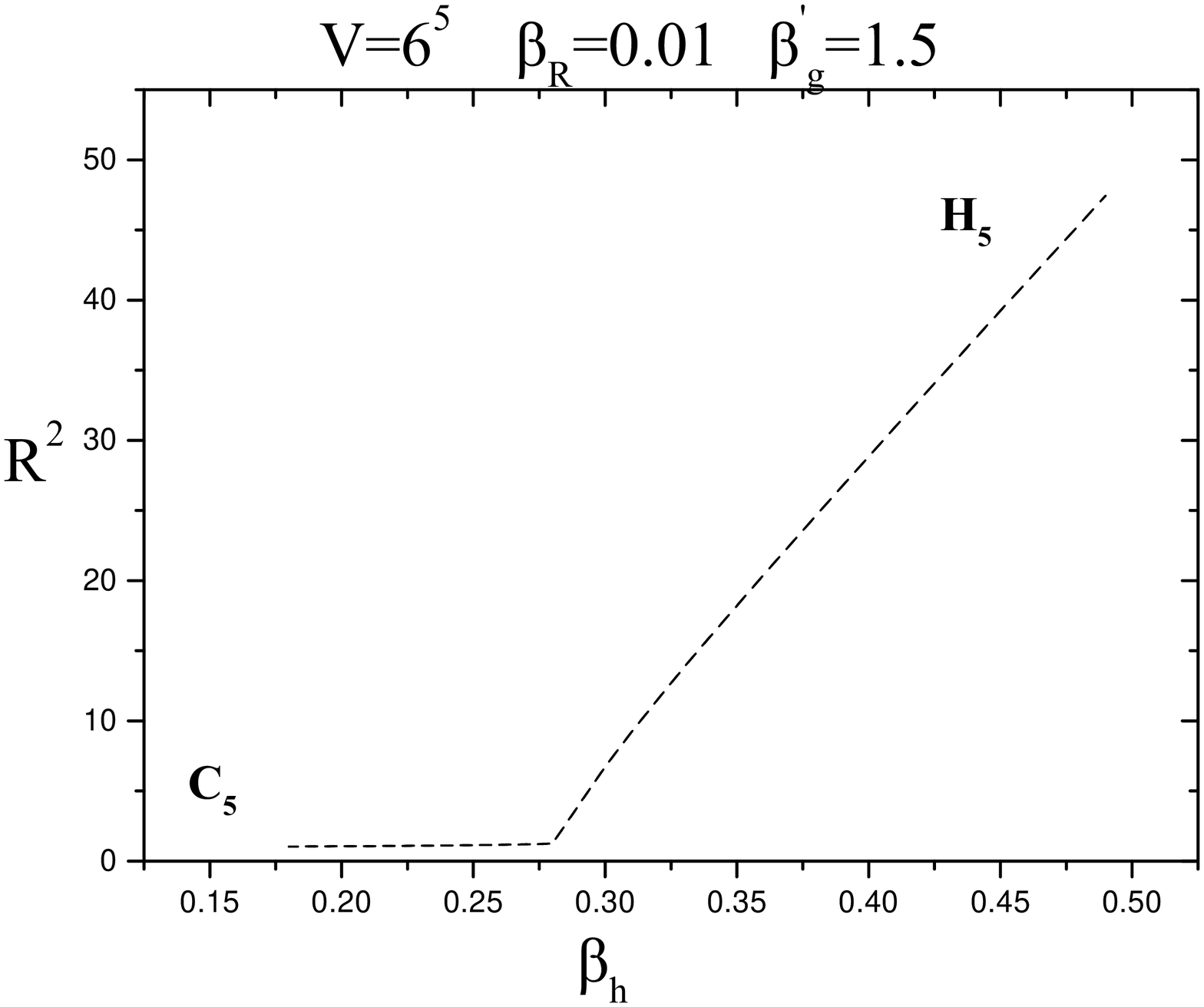}}
\caption{Hysteresis loop for  space--like and transverse--like link
(a) and $R^{2}$ (b), for  $\beta_{g}^{\prime}=1.5$.}
\label{Fig7}
\end{figure}

Finally, the transition between $C_{5}$ and $H_{5}$ is shown in
Figure \ref{Fig7}  where we set $\beta_{g}^{\prime}$ to 1.5 and let
$\beta_{h}$ run.  In Figure \ref{Fig7} (a) we show
the behaviour of the space--like and the
transverse--like link from which we can see that they both exhibit a gradual
increase as $\beta_{h}$ grows. In addition, the corresponding behaviour of
$R^2$ values in Figure \ref{Fig7} (b) implies the transition from
a five--dimensional Coulomb phase to a five--dimensional Higgs
phase.

\begin{figure}[!h]
\begin{center}
\includegraphics[scale=0.3]{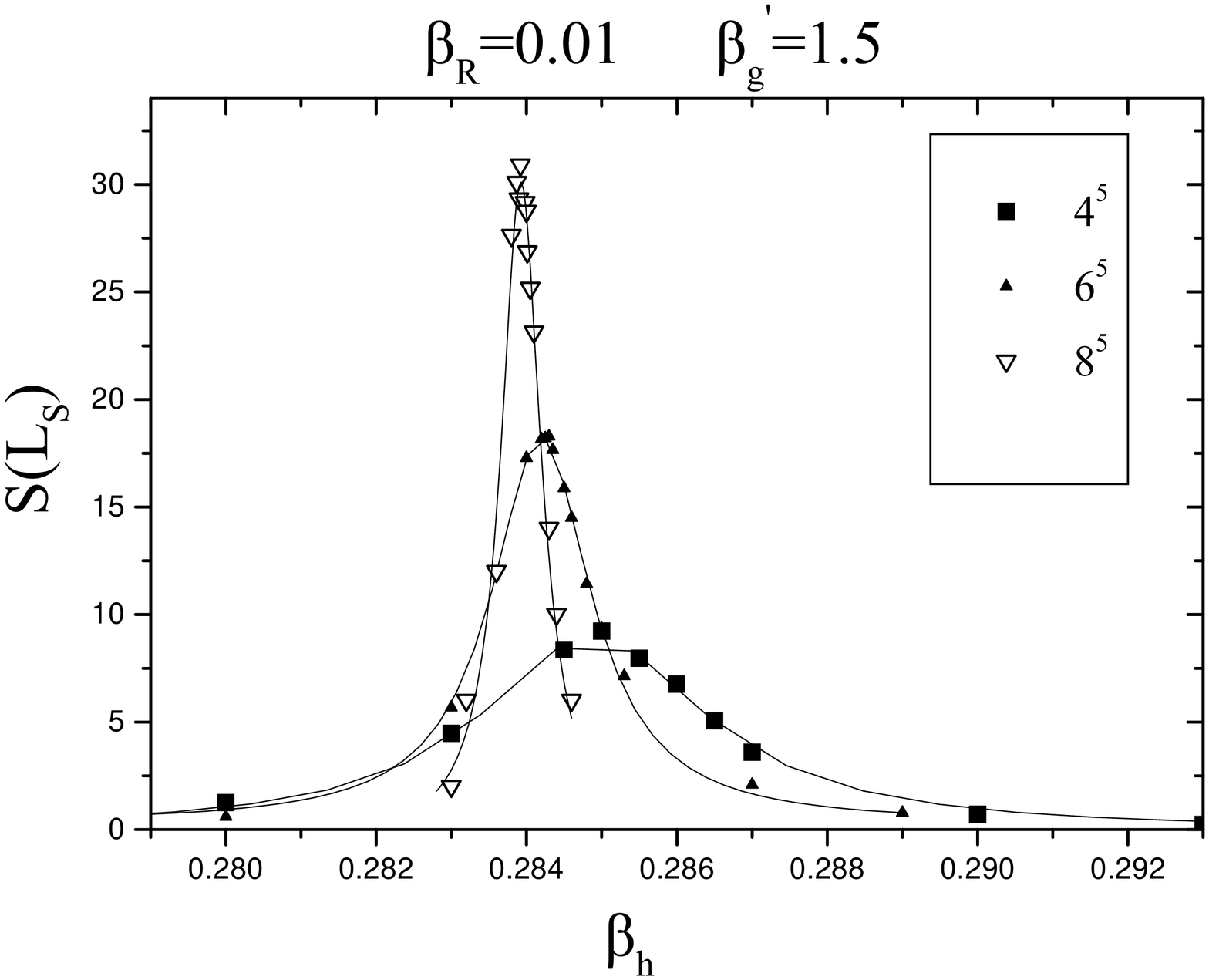}
\caption{Susceptibility $S(L_{S})$ for three values of lattice volume for $\beta_{g}^{\prime}=1.5
$ (the error--bars are smaller than the sizes of the symbols). }\label{Fig8}
\end{center}
\end{figure}

\begin{figure}[!h]
\subfigure[]{\includegraphics[scale=0.3]{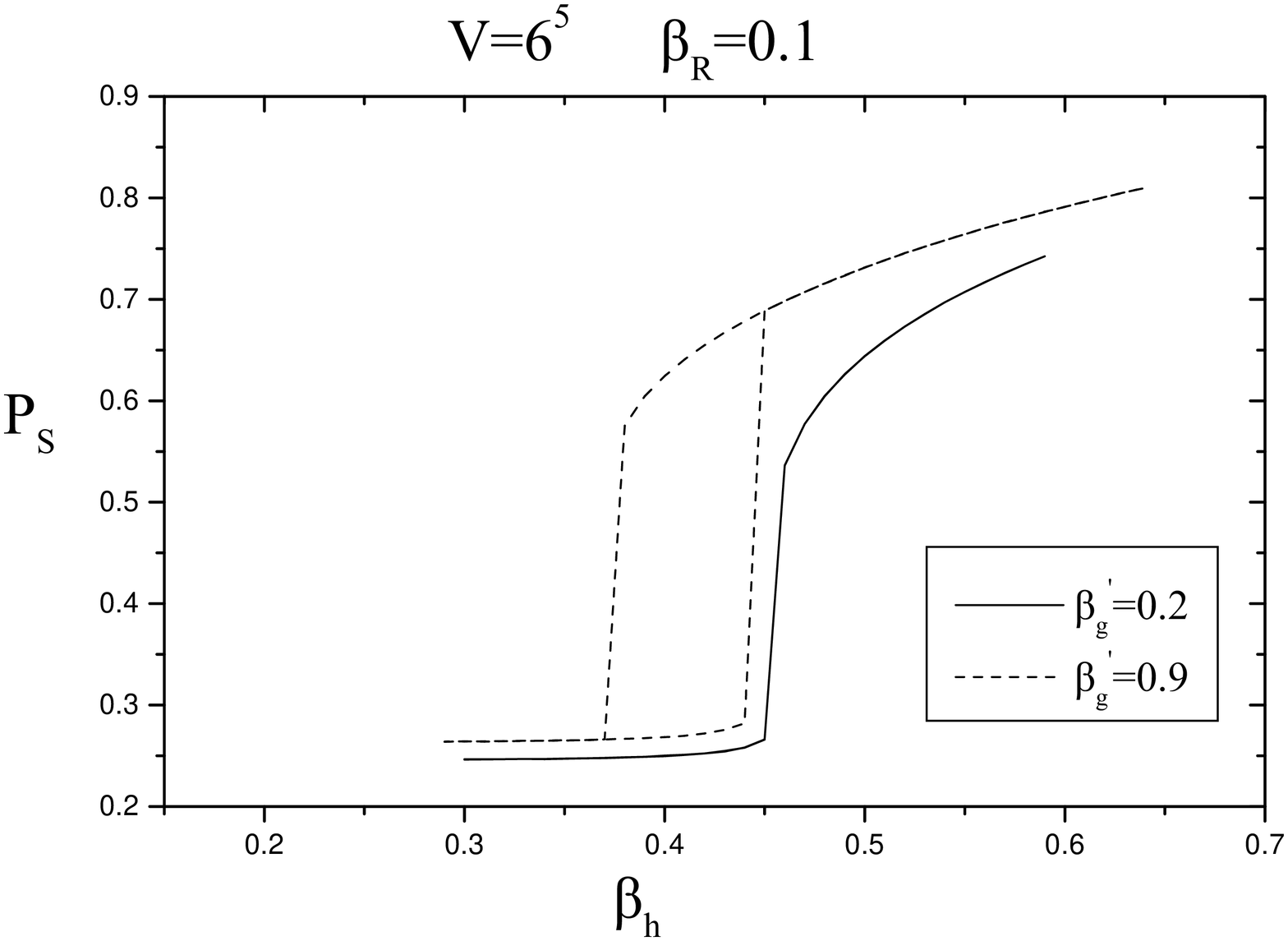}}
\subfigure[]{\includegraphics[scale=0.3]{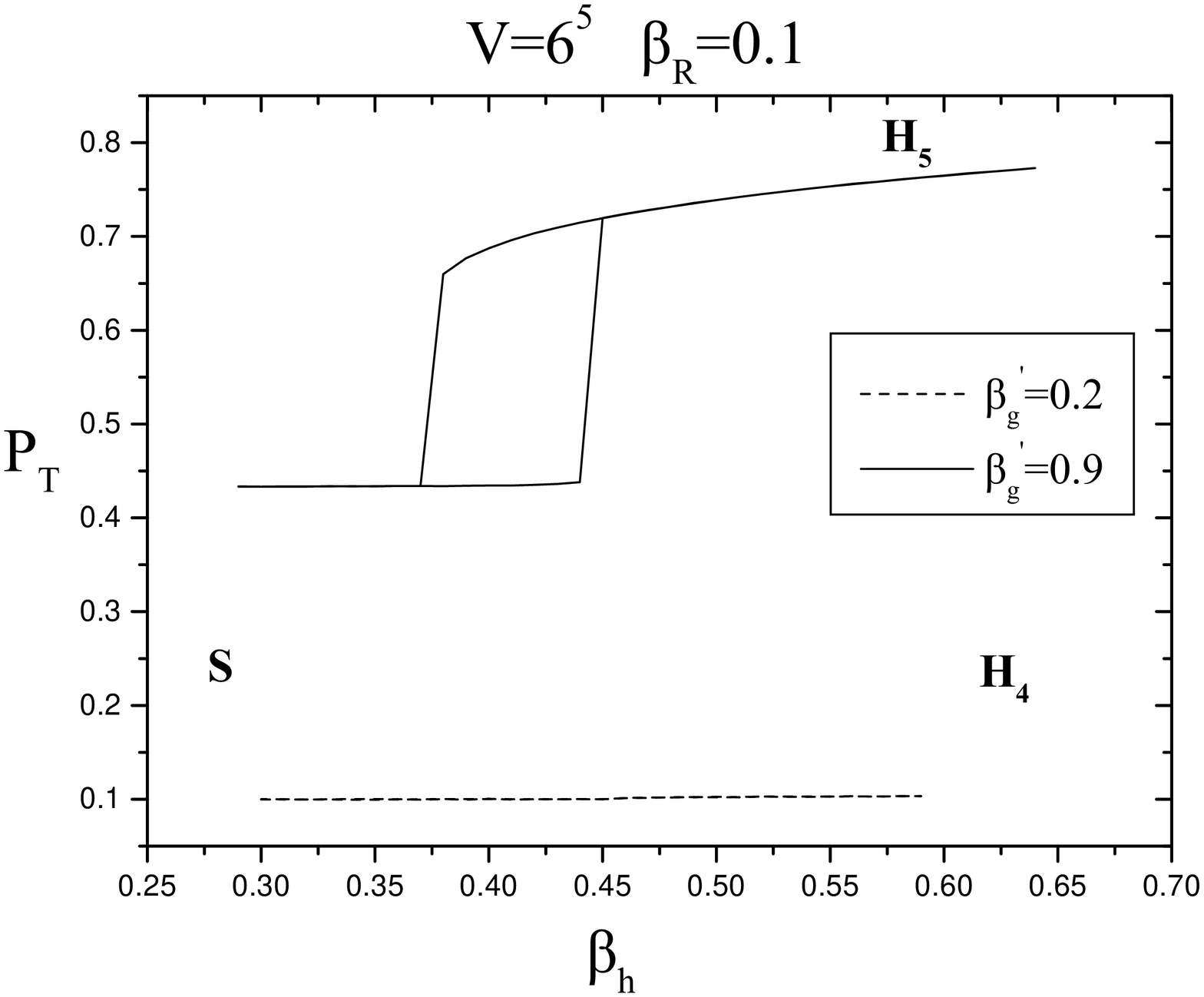}}
\caption{Hysteresis loop for  space--like (a) and transverse--like
plaquette (b)
for $\beta_{g}^{\prime}=0.2$ and $\beta^{\prime}_{g}=0.9$.}
\label{Fig9}
\end{figure}

\begin{figure}[!h]
\begin{center}
\includegraphics[scale=0.3]{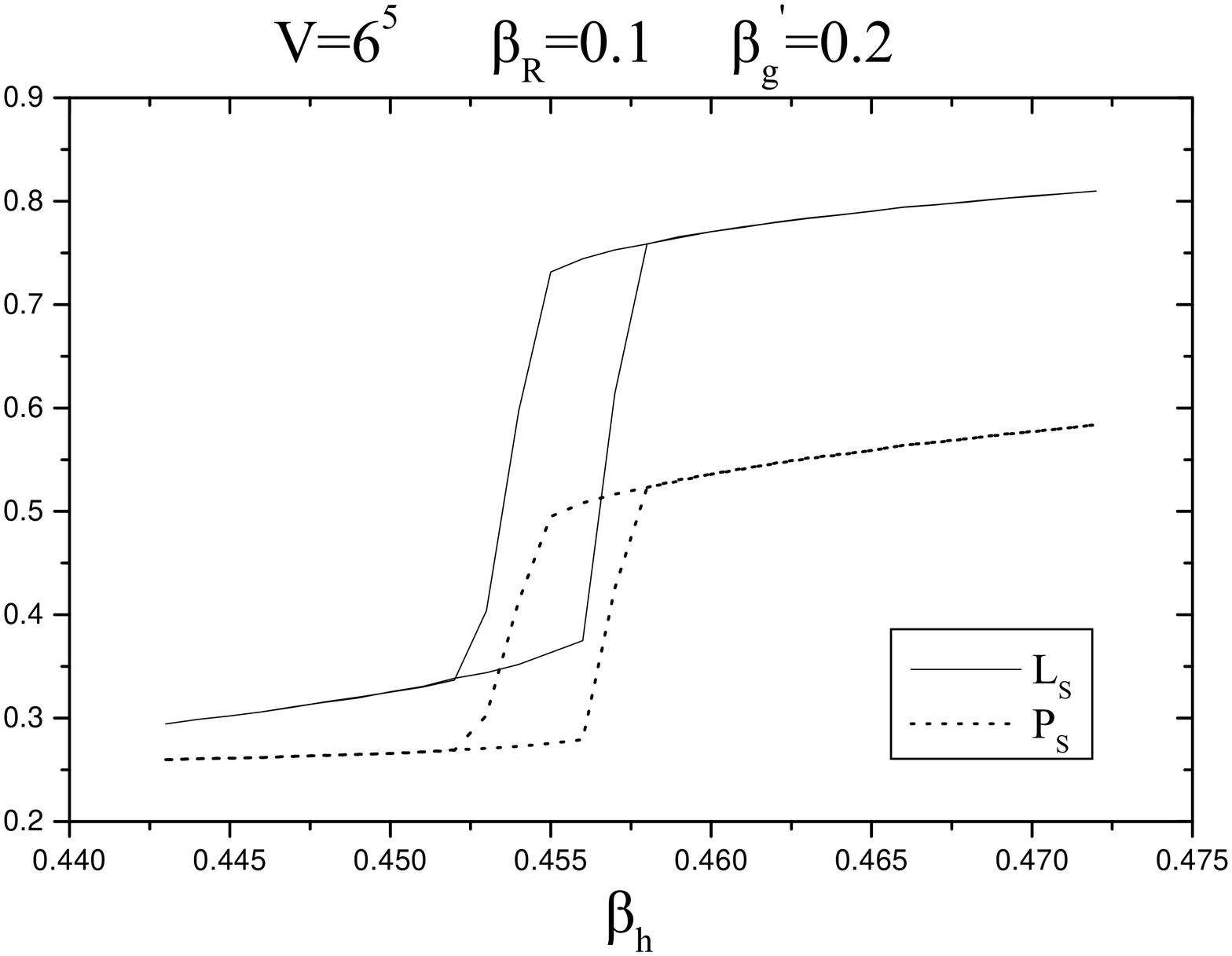}
\caption{Hysteresis loop for space--like link and plaquette for
$\beta_{g}^{\prime}=0.2$ in more detail, indicating a first order
phase transition. }\label{Fig10}
\end{center}
\end{figure}

The order of the phase transition
is not obvious and further study will be needed. To this end we
measure the susceptibilities for the $L_{S}$ order
parameter at the value $\beta_{g}^{\prime}=1.5$. The results shown in
Figure \ref{Fig8}
seem to be consistent with a second order phase transition as the maximum
values of susceptibility for each volume grows with some power of the
volume which is obviously much less than unity. \vspace{0.2cm}

Up to now, we
have given some examples of the  behaviour of the system for
$\beta_{R}=0.01$. In general, the phase structure for $\beta_{R}$
an order of magnitude larger than this , e.g. $\beta_{R}=0.1$,
is similar. However, we can elaborate on some points
concerning the structure of the layered phase which is formed in the
$S - H_{4}$ transition on one side and the
$S - H_{5}$ transition on the other.

In Figure \ref{Fig9} we can see an example for the transition from $S$ to
$H_{4}$ ($\beta_{g}^{\prime}=0.2$) and from $S$ to $H_{5}$
($\beta_{g}^{\prime}=0.9$), by giving the hysteresis loops for
$P_{S}$ and $P_{T}$. The difference of this figure
with Figure \ref{Fig3} consists in the ``weaker" transition from $S$ to
$H_{4}$ phase, since the loop is much smaller. A more detailed study
for the case
$\beta_{g}^{\prime}=0.2$ shows (Figure \ref{Fig10}) that a
clear hysteresis loop is formed, indicating a first order phase
transition, though seemingly weaker, than that of $\beta_{R}=0.01$
case. However, the trasverse--like order parameter $P_{T}$ does
not change at all as $\beta_{h}$  increases. This fact points out
that a layered phase has appeared and the layers are
decoupled from each other.

\subsection{Multi--layer Structure}
\begin{figure}[!h]
\subfigure[]{\includegraphics[scale=0.3]{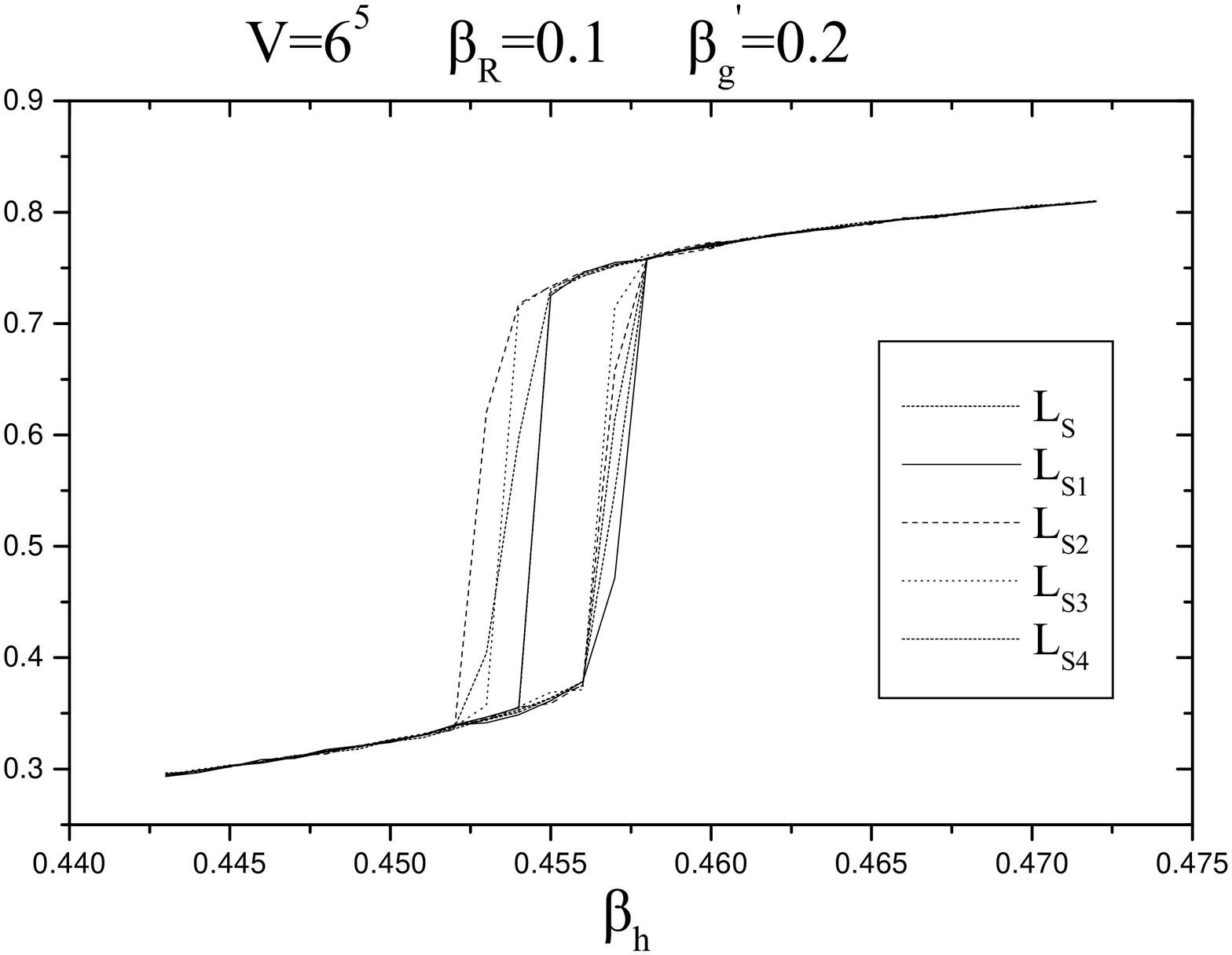}}
\subfigure[]{\includegraphics[scale=0.3]{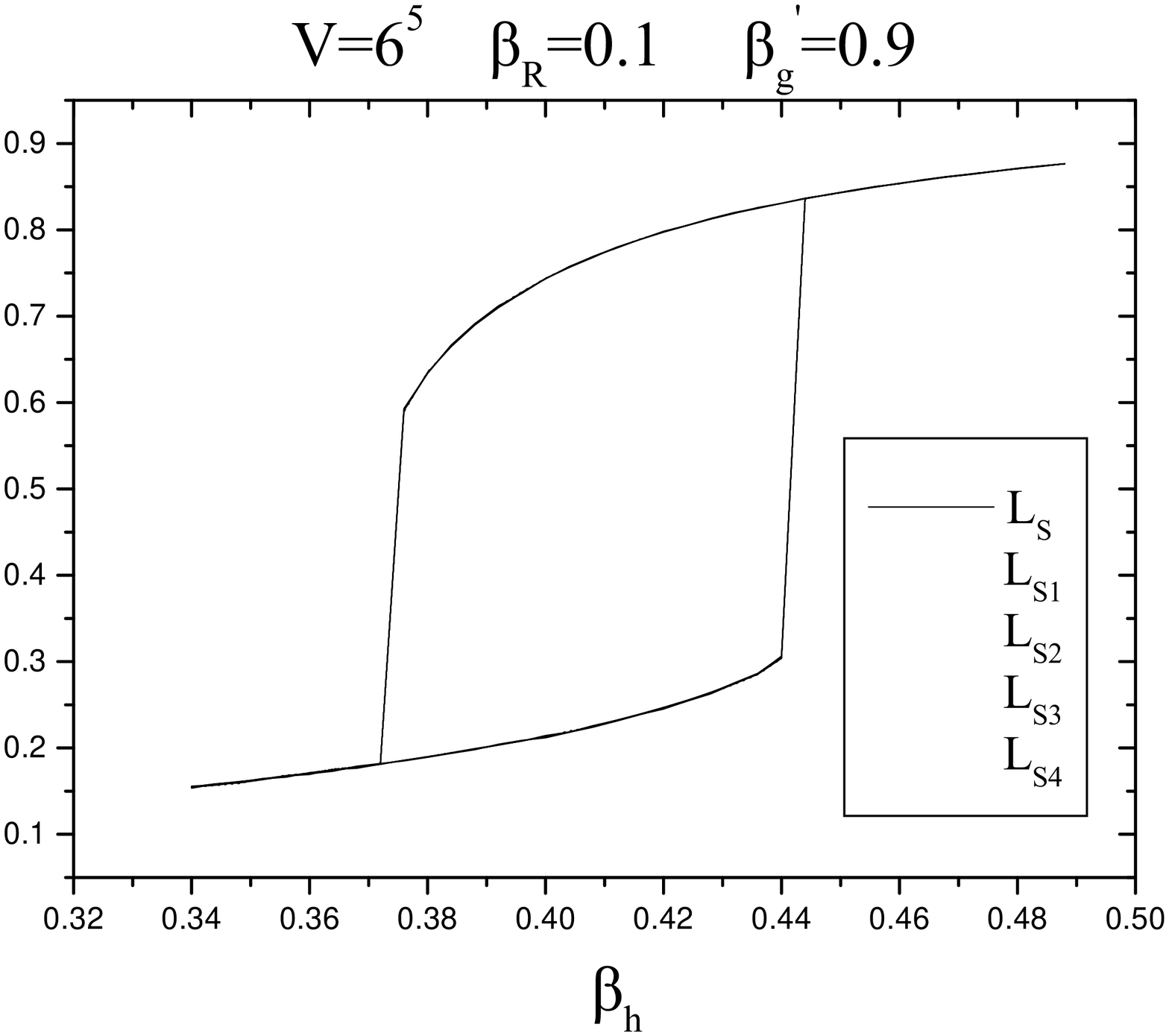}}
\caption{Hysteresis loop for $\beta_{g}^{\prime}=0.2$ (a) and
$\beta_{g}^{\prime}=0.9$ (b). We show the very different way
for the transition of $L_{S}$ calculated either on the
five dimensional volume and on four layers. In the figure on the
left the $S - H_{4}$ transition is presented where the layers show
a decoherent behaviour on the phase transition in contrast with
the $S - H_{5}$ phase transition on the right where the transition
for $L_{S}$ is identical for all layers, so it coincides with the
mean values over the 5D--volume.}
\label{Fig11}
\end{figure}

The next step is to study the behaviour of the layers one by one
as the system moves through the phase transition. This is shown in
Figures \ref{Fig11} (a) and (b) which correspond to $S - H_{4}$ and $S - H_{5}$
phase transitions respectively. In the Figure \ref{Fig11} (a), corresponding
to the $S - H_{4}$ case,  one
may easily see
a ``non--sychronised" transition exhibited by $L_{S}$ defined on
each space--like
volume (layer) in contrast with Figure \ref{Fig11} (b),
in which the corresponding
order parameters  indicate the phase transition simultaneously
(Obviously, in figure (b) the hysteresis loops
formed by the $L_{S}$ defined on each layer can not be distinguished from
the corresponding hysteresis loop due to $L_{S}$ defined on the volume.)
This specific behaviour of hysteresis loops, may
actually  serve as  a ``criterion" to characterise
the layered phase.

\begin{figure}[!h]
\begin{center}
\includegraphics[scale=0.3]{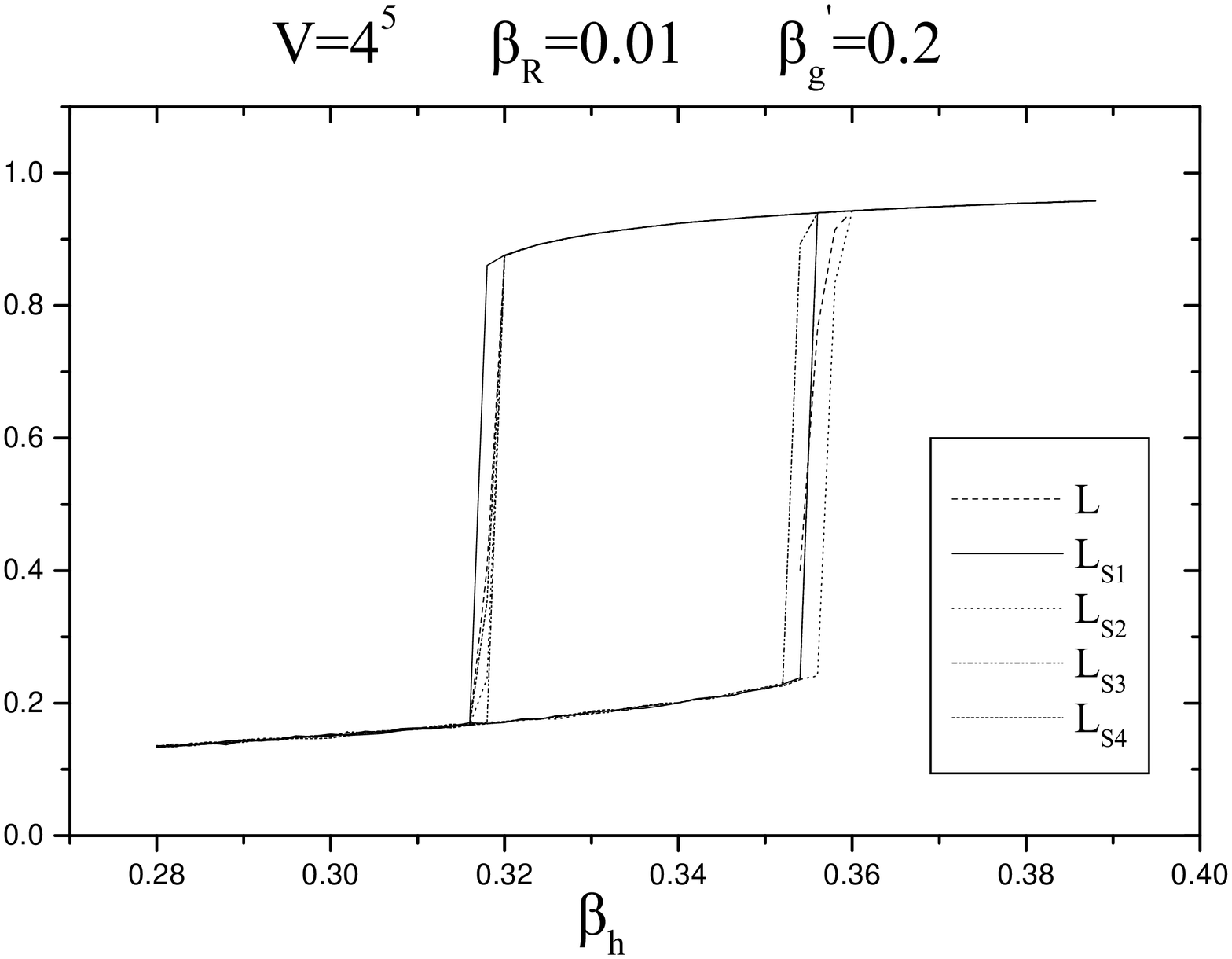}
\caption{Hysteresis loop for $L_{S}$ showing the way that the layers
decouple for $\beta_{R}=0.01$ at $\beta_{g}^{\prime}=0.2$. }\label{Fig12}
\end{center}
\end{figure}

We also reach the same conclusion by considering  the
$\beta_{R}=0.01$, $\beta_{g}^{\prime}=0.2$ case, for a $4^5$ lattice
in more detail. The
result, shown in Figure \ref{Fig12}, leads to the same conclusion:
The very existence of the layered
phase (and consequently  of confinement in the fifth dimension)
shows up in the ``incoherent'' behaviour of the space--like
volumes (layers) as the phase transition takes place.

\begin{figure}[!h]
\subfigure[]{\includegraphics[scale=0.3]{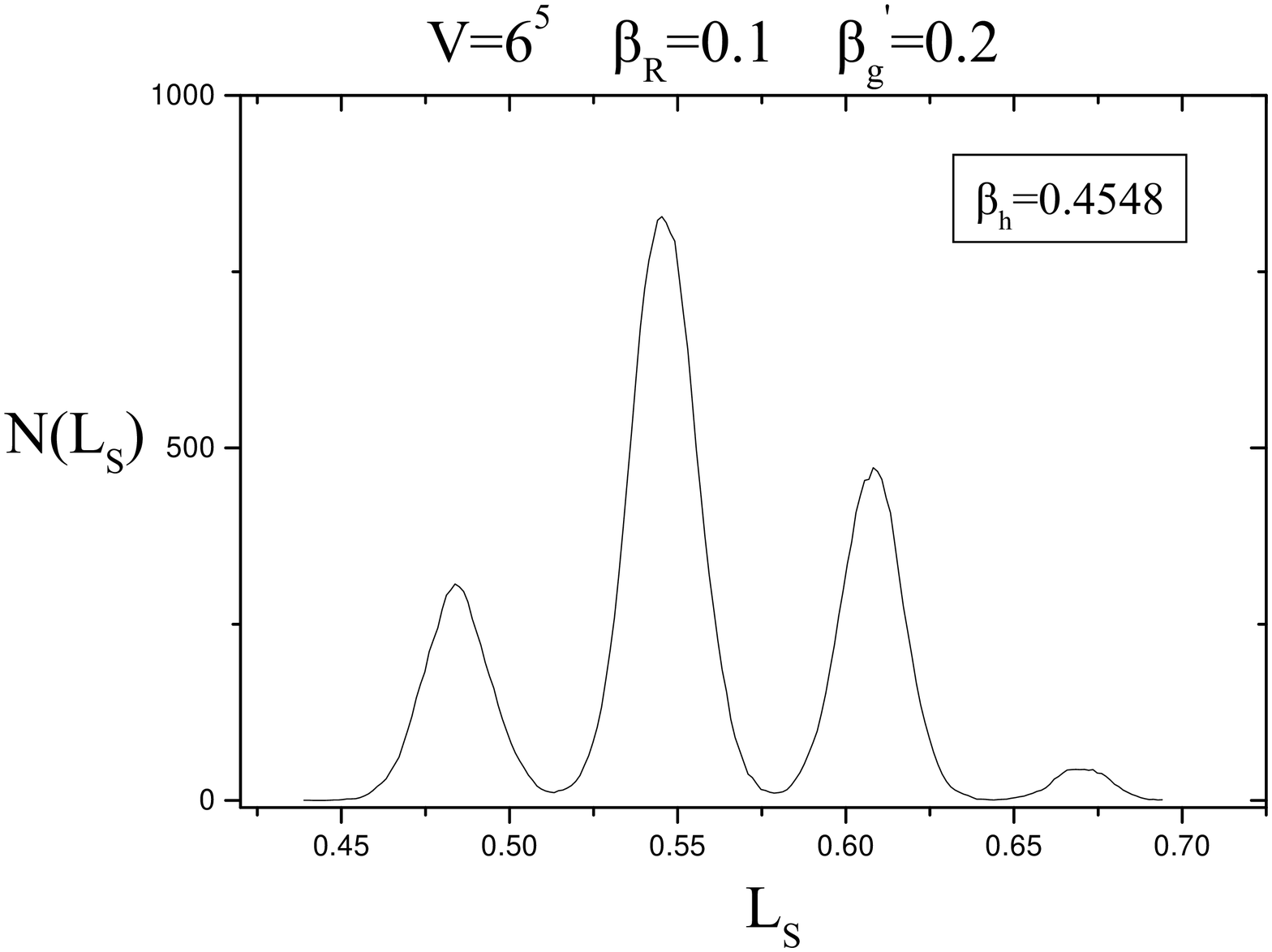}}
\subfigure[]{\includegraphics[scale=0.3]{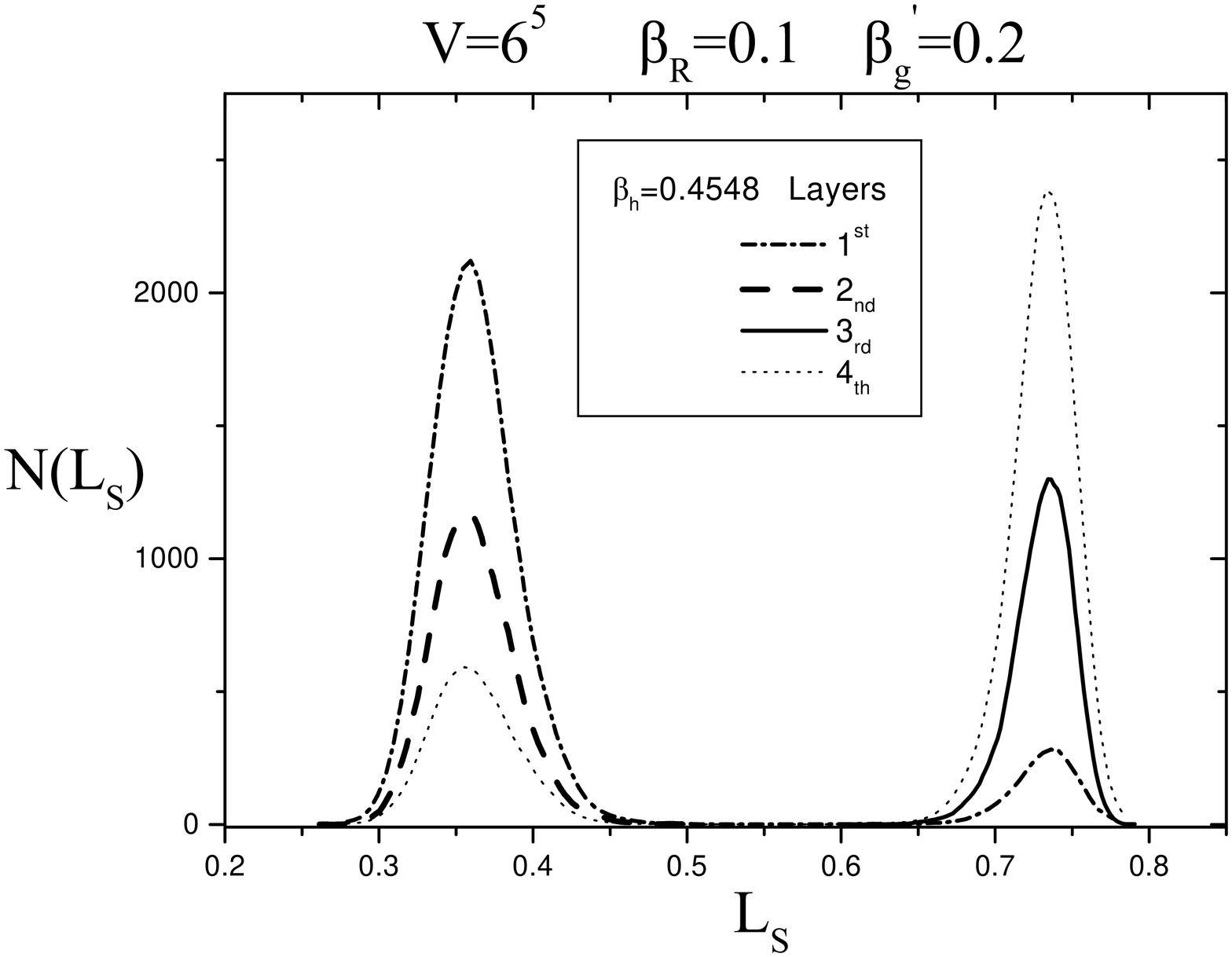}}
\caption{Distribution for $L_{S}$ on the volume (a) and on each
layer (only four of the six are shown in the figure)(b) for
$6^5$ lattice  in the critical region for the $S - H_{4}$
phase transition.}
\label{Fig13}
\end{figure}

This  behaviour can be  further confirmed by long runs
in the transition region. It is to be expected
that the $S - H_{4}$ phase transition is of first order. Therefore,
we would expect a two peak signal in the order parameter
disrtibution at equilibrium. Nevertheless the situation, shown in
Figure  \ref{Fig13} (a) concerning the distribution for the order parameter
$L_{S}$  for a value of $\beta_{h}$
near  the transition region  for $V=6^5$
is by no means what one would expect normally. The multipeak structure
seems rather strange. {It should be noticed that the same occurs for
the order parameter $P_{S}$ too. }However, the study of the same order
parameter defined on each space--like volume is more illuminating. For
example in Figure \ref{Fig13} (b), we can see
the distribution of $L_{S}$
values on each layer.
We show four out of six distributions of $L_{S}$ corresponding to the
four  space--like layers within the five--dimensional volume.
The  distributions which are produced appear quite usual and they show
that at the same time one layer is in the strong phase (called $2_{nd}$ in
the figure), another has already
passed to the broken phase ($3_{rd}$) and others produce a two peak signal result.
The result is that the strange picture of the distribution formed in Figure
\ref{Fig13} (a) is  resolved if we analyze the behaviour of
the system on each layer as the system undergoes the phase
transition.

This result is found  for all of the  volume sizes (e.g. $4^5, 6^5, 8^5$)
which we have worked on. Although it is consistent with a first
order phase transition it lends support to the view of
a  {\it decoherent} behaviour for every four--dimensional
volume (layer) in the transition region between five--dimensonal
strong phase and the four--dimensional layered Higgs phase.
This certain behaviour describes a dynamical decoupling of the
layers and provides a possible mechanism for localization of
the fields on the layers.

\subsection{Mean Field Approach}
The Mean Field Approach provides a point of view complementary
to  Monte Carlo sumulations. Although it is, by construction, blind to
spatial fluctuations, and cannot, therefore, see the multi-layer structure,
it does not suffer from the finite volume effects, that limit Monte Carlo
simulations. Furthermore, it is expected to be a reliable guide to the phase diagram,  the higher the dimension of the system under study.

Thus, in this work we use the Mean Field
analysis  (i) to show that the small value of
$\beta^{\prime}_{h}$ we have chosen is really a suitable one to
reveal the layered Higgs phase and (ii) to provide evidence
that as Higgs self--coupling takes larger values --up to
$\beta_{R}=0.2$-- the phase transition from the Strong to the $H_{4}$
phase may become  second order.

We start with the action (\ref{compactaction}).
We  fix the gauge by imposing $U_{\hat 4}(x)=I$ and use the
translation-invariant {\em
Ansatz}~\cite{funiel, stam, dim2},
$U_{\hat \mu}(x)=v,\,1 \le \mu \le 3$; $U_{\hat 5}(x)=v'$. We also
introduce the variables for the Higgs field,
\begin{equation}
\label{Higgs}
\phi(x)=\rho(x)e^{{\mathrm i}\chi(x)}
\end{equation}
and have also assumed a translationally invariant {\em Ansatz},
$\rho=\rho(x)$, $\chi=\chi(x)$.
The free energy, which should be minimized to get the mean field
solution, reads:
$$
F=-\beta_g 3 v_a^4-\beta_g 3 v_a^2
$$
$$
-\beta_g^\prime 3 v_a^2 v_{a'}^{\prime 2}-\beta_g^\prime v_{a'}^{\prime 2}
$$
$$
-(3 \beta_h v_a +\beta_h +\beta_h^\prime v_{a'}^{\prime}) \rho^2 v_\chi^2
$$
$$
+(1-2 \beta_R) \rho^2+\beta_R \rho^4-\frac{1}{2} log[\rho^2]
$$
\be
+3 a v_a-3 log[I_0(a)]+a^\prime v_{a'}^\prime-log[I_0(a^\prime)]
+\chi v_\chi-log[I_0(\chi)]
\label{fren}
\ee

\noindent The parameters $a$, $a'$ and $\chi$ are conjugate to $v_a$,
$v'_{a^\prime}$ and $v_\chi$ respectively. \\ We study the phase
structure  using the order parameters defined in Section 2.

Our Monte Carlo analysis has been performed setting
$\beta_{h}^{\prime}=0.001$. It is of interest to have an idea what
happens when $\beta_{h}^{\prime}$ takes other values.

\begin{figure}[!h]
\begin{center}
\subfigure[]{\includegraphics[scale=0.3]{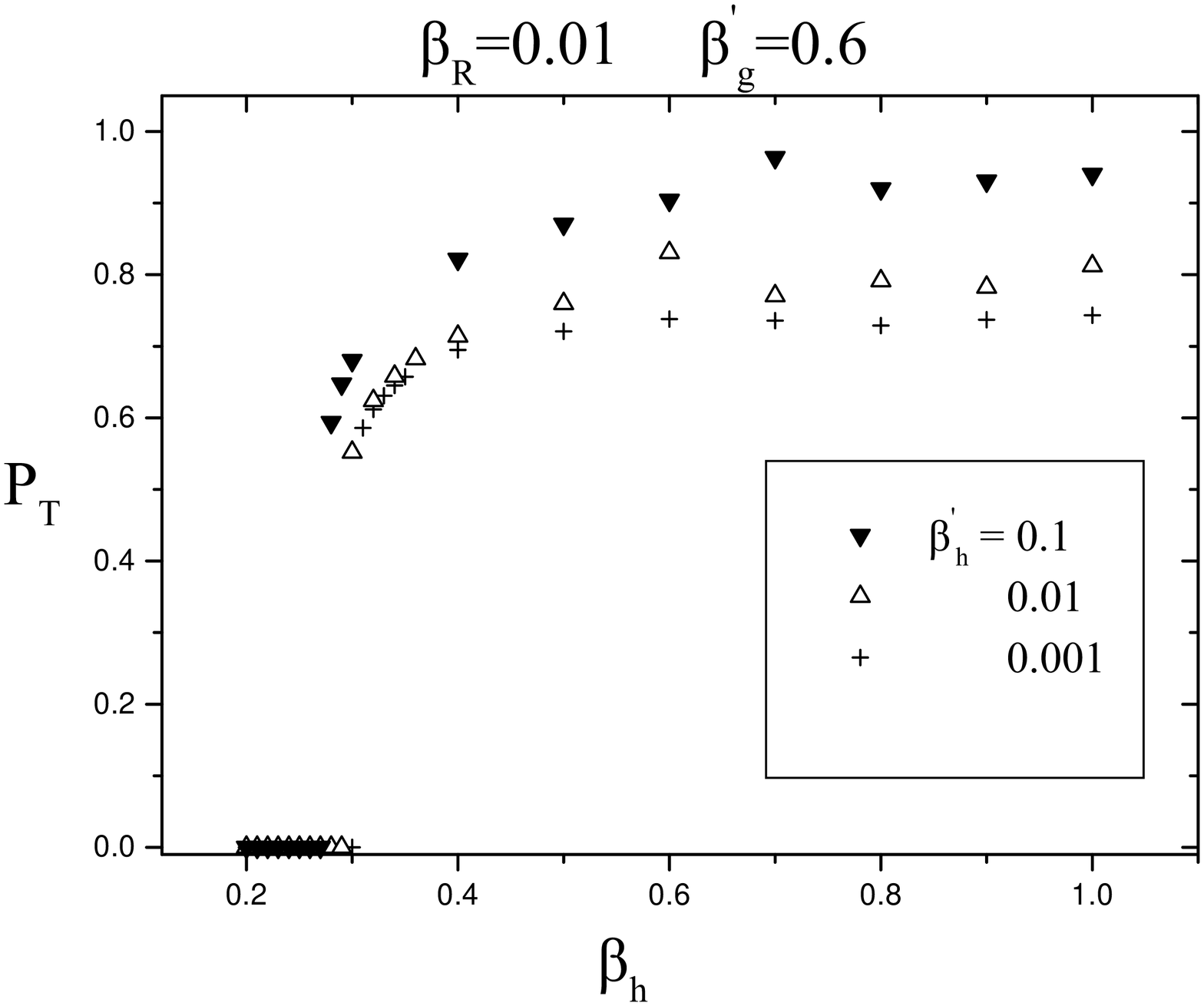}}
\subfigure[]{\includegraphics[scale=0.3]{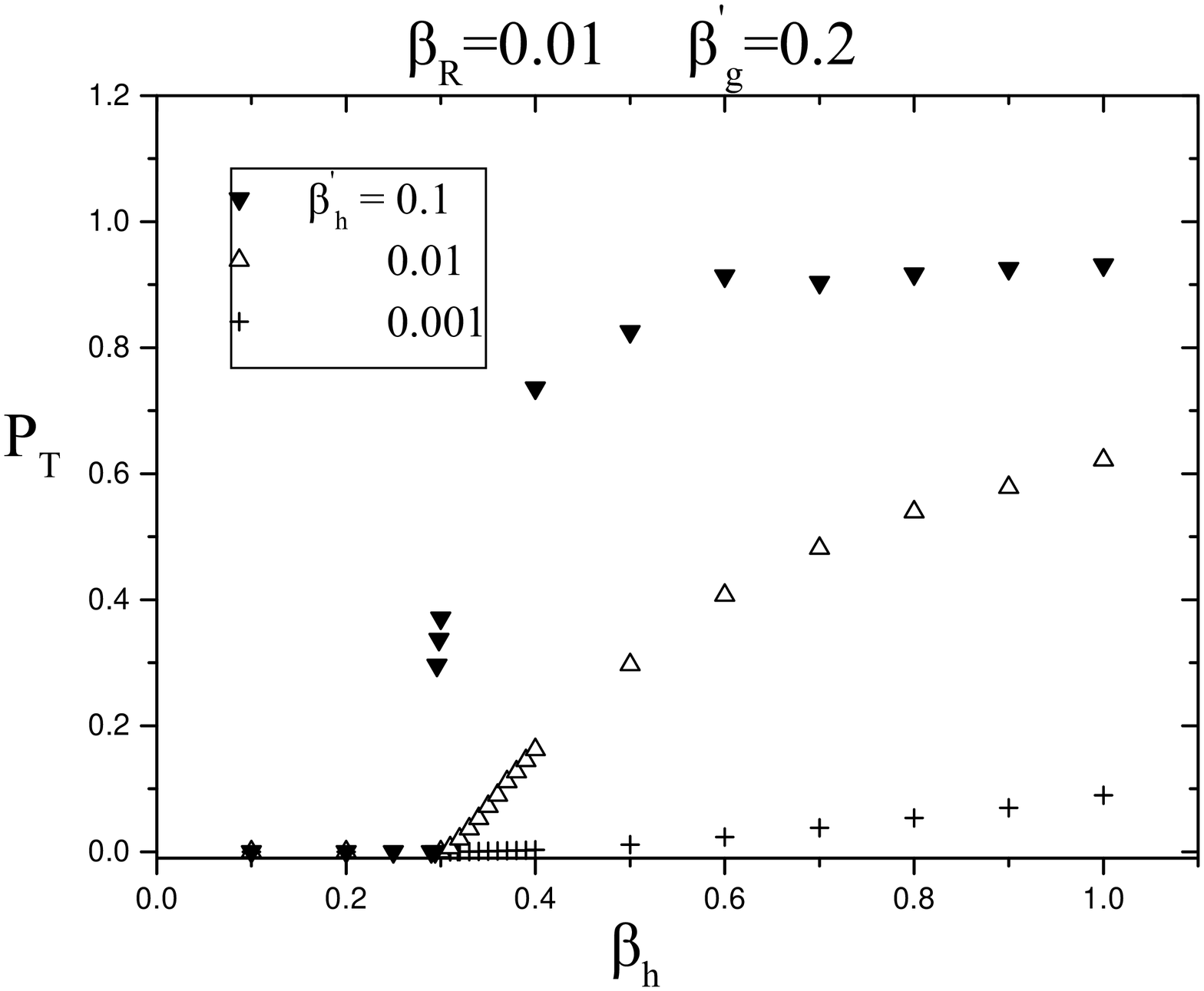}}
\caption{We show the behaviour of $P_{T}$
as the value of $\beta_{h}^{\prime}$ grows. The transition
from Strong phase to the $H_{5}$ phase does not seem to change (a), in
contrast with the  one to from Strong to $H_{4}$
which is washed away (there is a radical change to the value of
$P_{T}$ at a critical point of $\beta_{h}$) leading thus to the
existence of only a 5D--Higgs phase.}
\label{Fig14}
\end{center}
\end{figure}
In Figure \ref{Fig14} we depict the behaviour of $P_{T}$ setting
$\beta_{R}=0.01$ and for the two
cases corresponding to the transitions from the Strong phase to $H_{5}$ and
$H_{4}$. The corresponding values for $\beta^{\prime}_{g}$ are 0.6
and 0.2 respectively (see, also, Figure \ref{Fig1}). From Figure
\ref{Fig14} (a) we can see that as $\beta^{\prime}_{h}$ increases
the transition behaviour to $H_{5}$ is fairly the same. On the
contrary, Figure
\ref{Fig14} (b)  shows that the increase in  $\beta^{\prime}_{h}$
leads to  a big increase in the value of $P_{T}$
which becomes compatible with the value characterizing the $H_{5}$
phase. This provides an indication
that at some value, $\beta^{\prime}_{h} \sim 0.1$, the
$H_{4}$ phase transforms to  $H_{5}$: as $\beta^{\prime}_{h}$ increases $H_{5}$
extends,  covering the region occupied before by $H_{4}$.

\begin{figure}[!h]
\begin{center}
\includegraphics[scale=0.3]{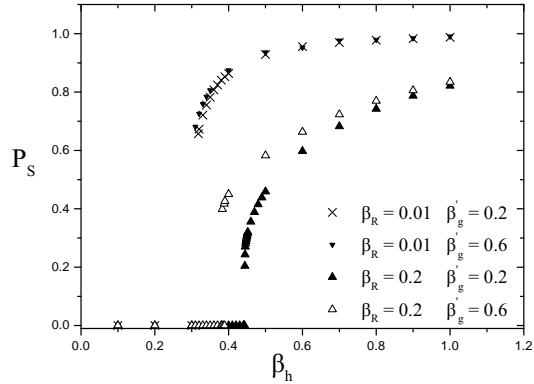}
\caption{As the value of $\beta_{R}$ increases the $S - H_{4}$
phase transition becomes weaker, as expected, while the
$S - H_{5}$  transition remains first order.}\label{Fig15}
\end{center}
\end{figure}

Figure \ref{Fig15} provides evidence of the
expected weakening of the Strong -- $H_{4}$ phase transition as
$\beta_{R}$ increases. It reveals that, although for
$\beta_{R}=0.01$, $P_{S}$ exhibits the same behaviour for
$H_{5}$ and $H_{4}$ phases, for $\beta_{R}=0.2$ the picture
changes substantially. The $S - H_{4}$ phase transition provides a
sign of a smoother phase transition leading probably to second
order. Mone Carlo simulations in this region of parameter space are required
to complement this evidence.

\section{Conclusions}
In this paper, we have shown,
using Monte--Carlo methods,  that a layered Higgs  phase actually exists
in the phase diagram of the strongly coupled five--dimensional Abelian
Higgs model and that it emerges from the confining bulk phase through
a first order phase transition. In fact we find the existence of multi-layers, as each brane passes from the
confining phase to the Higgs phase independently of the others.

Using mean field techniques we performed a scan of the phase diagram as
the Higgs couplings changed and found evidence that, as the Higgs self-coupling increases, the emergence of the Higgs layers from the confining bulk phase
softens and may become second order, leading to new continuum theories.
Further work will clarify this issue.

It is worth stressing that the layers of our model are, indeed, 3-branes.
In string theory one expects symmetry enhancement when branes coincide and
the question arises, whether such an effect could be visible within our field
theory context. This effect is the manifestation of new, non-perturbative,
degrees of freedom. In our context this would mean introducing magnetic
monopoles, that would promote these 3-branes into true D-3-branes. One way of
achieving this could be using twisted boundary conditions, along the lines
of  ref.~\cite{arroyo-altes}.

This raises the question of what aspects of our study may also be of relevance
to Yang-Mills theories~\cite{china,rab,ejiri}.
 As is well known, these are confining in less than
five dimensions, so do not seem to admit four dimensional layered phases.
The simplest example would, thus, be an anisotropic $SU(2)$ theory in six
dimensions and in this case  the layers form an
unnatural five--dimensional Coulomb phase . The interesting point
would be if  the SU(2) Higgs model with  extra dimensions could reveal
a layered Higgs phase in four dimensions
which bears a closer resemblance to the SM.
Monte Carlo simulations of such theories directly are,
unfortunately,
inconclusive with current technology--though the use of massively parallel
clusters may offer some hope for the not too distant future. Using suitably
reduced models, on the other hand, such as the one studied here, could provide
some useful information on their structure.

\vspace{1cm}

\noindent{\bf Acknowledgements} \vspace{0.2cm}

\noindent The authors are grateful to G. Koutsoumbas for extensive discussions
on the subject during the preparation of the paper and for reading the
manuscript.
P.D. and K.F. would like to thank A. Kehagias for useful
discussions on related topics. Also, P.D. and  K.F.
acknowledge support from the TMR project ``Finite Temperature
Phase Transitions in Particle Physics", EU contract
FMRX-CT97-0122.

\end{document}